\documentclass{article}
\usepackage{spconf,amsmath,graphicx,hyperref}
\usepackage{amssymb,booktabs,array,multirow,xcolor}
\hypersetup{pdftitle={AccentCanvas: Activity-Guided Bidirectional Accent Conversion}}

\newcommand{\our}{AccentCanvas}

\title{AccentCanvas: Activity-Guided Bidirectional Accent Conversion}
\name{Jaehun Kim, Kang Zhang, Sung Woo Kim, Junmo Kim, Joon Son Chung}
\address{Korea Advanced Institute of Science and Technology, South Korea}

\begin{document}
\maketitle
\begin{abstract}
Accent conversion for dubbing must change pronunciation while preserving content, voice, and source timing.
Normalization alone cannot produce requested non-native accents, and matching total duration does not ensure local speech/pause alignment.
We present \our{}, a shared discrete diffusion model for normalization and L1-to-L2 conversion, trained on cross-paired shared-prompt recordings and synthetic pairs.
Frame-level speech activity and repetition-preserving units guide local timing through synthesis.
Across both tasks, \our{} achieves the highest accent conversion performance (Accent Arrival) and mean listener-rated accent similarity among compared systems, with over 73\% lower speech-boundary error than the best baseline for each task.
These gains accompany higher L1-to-L2 WER than TokAN and Vevo-Style, highlighting a tradeoff between requested-accent resemblance and ASR-based content recognition.
\url{https://smokedindia.github.io/accentcanvas/}
\end{abstract}

\section{Introduction}
\label{sec:intro}
Accent conversion changes the pronunciation of speech while preserving its linguistic content and speaker identity.
This capability supports applications in communication, language learning, and media production~\cite{jin2023multiple}.
In speech dubbing, converted speech must also fit the timing of the original.
Much recent work focuses on accent normalization, which converts L2-accented English toward native American English.
Recent systems such as TokAN~\cite{tokan}, CosyAccent~\cite{cosyaccent}, and DLM-AN~\cite{dlman} investigate this through discrete speech representations and generative models, addressing tradeoffs among accent change, intelligibility, and speech quality.

The focus on normalization raises a complementary question: can a conversion model also transform native English toward a requested L2-English accent?
For example, an English dub may preserve a Korean actor's voice while converting the dialogue from an American accent to the Korean-accented English familiar to the actor or production, which requires conversion in the inverse direction.
Several methods have explored this broader setting.
RWA~\cite{rwa} performs many-to-many conversion through phoneme remapping and duration adjustment, while the system requires accent-specific pronunciation resource.
PSDN~\cite{jia2023psdn} disentangles accent from content and demonstrates foreign-to-British and American-to-Indian conversion, although its experiments train for a designated target only.
More recently, Bagat et al.~\cite{bagat} adapt TokAN for reference-conditioned L1-to-L2 conversion.
Their evaluation focuses on improving speech recognition, leaving perceptual and per-accent evaluation to future work.

Limited accented speech data remains a central challenge in learning accent conversion.
To mitigate this, TokAN~\cite{tokan} synthesizes native target speech, CosyAccent~\cite{cosyaccent} synthesizes accented sources, and DLM-AN~\cite{dlman} uses both source- and target-synthesized pairs using text-to-speech (TTS) models.
Synthetic supervision also includes transliteration-based multi-accent pairs~\cite{inoue2025macst} and native-TTS targets~\cite{nguyen2025distillation}.
We propose expanding the use of real parallel accent-conversion data through cross-pairing.
Exploiting the shared prompts in CMU ARCTIC and L2-ARCTIC~\cite{arctic,l2arctic}, we construct content-matched source--target pairs across accent groups.
Recordings serve as both sources and targets across conversion relationships, expanding training supervision without collecting additional speech.

We then introduce \our{}, a shared discrete diffusion model that generates accent-converted speech based on the requested accent.
The model predicts the target speech unit conditioned on the source speech features via cross-attention, and the controllable-length initialization enables generation with varying speech durations.
To generate converted speech with speech--pause synchronization with the source, we train the model with frame--level activity guidance as an explicit condition.

\begin{figure*}[!t]
\centering
\includegraphics[width=\textwidth,pagebox=cropbox]{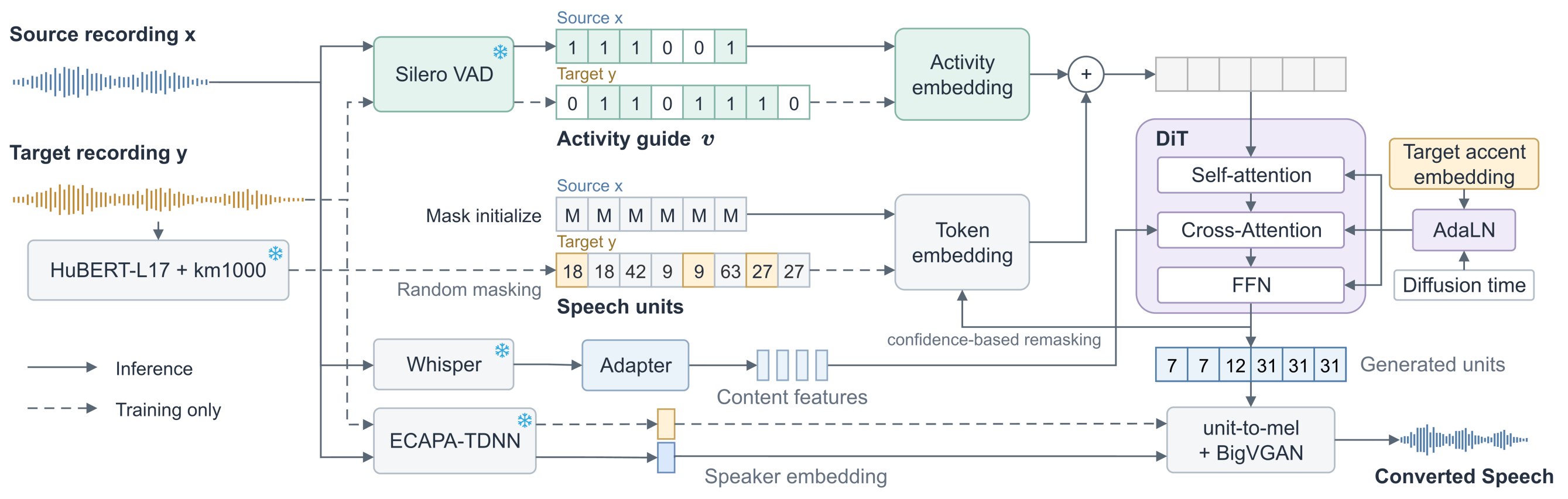}
\caption{
Overview of \our{}.
One model performs both normalization and L1-to-L2 conversion, and requires no text guidance during training.
Speech units are utilized with no de-duplication.
During inference, masked speech unit length is initialized to match the source speech duration.
Generated units are rendered with a source speaker reference. 
}
\label{fig:architecture}
\end{figure*}

We evaluate speech converted in both directions and demonstrate that \our{} generates both L1 and L2 accents with high naturalness.
The ablation results show that existing synthetic-only training can yield intelligibility; however, it fails to generate L2 accents in the converted speech.
The proposed cross-pairing achieves a lower WER than fixed one-to-one pairing while maintaining accent performance.
Our selected recipe combines cross-paired real recordings with a smaller synthetic contribution and achieves lower WER than the evaluated real-only recipes.
Activity conditioning effectively reduces speech-boundary error, supporting synchronized conversion suitable for dubbing applications.

\section{Method}
\label{sec:method}
\our{} converts a source utterance to a requested accent, and uses speech activity to guide the output timing.
As illustrated in Fig.~\ref{fig:architecture}, a discrete diffusion model predicts speech units from source-content features, a target accent label, and a frame-level speech activity sequence.
A pretrained renderer then synthesizes an audible waveform using the source-speaker reference.
\vspace{-0.1cm}
\subsection{Real speech dataset cross-pairing}

We expand the scope of leveraging accented speech datasets by cross-pairing samples from CMU ARCTIC and L2-ARCTIC with shared prompts.
For each prompt, we vary ordered source–target pairings across speakers and accent groups within the training partitions, yielding normalization, L1-to-L2, and cross-L2 pairs.
This maximizes the use of natural accent recordings across conversions without requiring additional data, exposing the training model to a wider range of conversion scenarios.

We train the model on these natural pairs together with a smaller contribution of CosyAccent synthetic pairs~\cite{cosyaccent}.
This mixture was deliberately designed to capture both the accent evidence from the natural recordings and content coverage from the synthetic pairs.
Section~\ref{sec:results} compares the resulting training-data recipes.
Note that, although the training \our{} includes TTS-synthesized data, the proposed model architecture requires no transcript information, enabling acceptance of broader accent datasets without text labels.
\vspace{-0.1cm}
\subsection{Conditional discrete diffusion}

We predict target speech units with a discrete diffusion model conditioned on source content, a requested accent, and speech activity~\cite{mdlm}.
A frozen speech encoder~\cite{whisper} extracts source features, which are projected and supplied through cross-attention.
Requested-accent and diffusion-time embeddings modulate the transformer blocks.

During training, speech units are extracted from the target speech, and valid positions are masked independently with probability $t_b=\max(U_b,0.001)$, where $U_b\sim\mathcal U[0,1)$, ensuring at least one masked position per $b$ sample.
We minimize
\begin{equation}
\mathcal L = -\frac{\displaystyle\sum_{b}\sum_{j\in\mathcal M_b}
 t_b^{-1}\log p_\theta(z_{b,j}\mid\mathcal C_b)}
 {\displaystyle\sum_b|\mathcal M_b|},
\end{equation}
where $\mathcal M_b$ is the masked position set and $z_{b,j}$ is the original unit.
The condition $\mathcal C_b$ comprises the masked units, source features, target accent, speech activity, and diffusion time.
During inference, the model iteratively unmasks a fully masked sequence based on prediction confidence~\cite{maskgit}.
The pretrained renderer synthesizes a waveform from the units, run lengths, and a source-speaker reference, following~\cite{tokan}.

\begin{table*}[t]
\centering
\fontsize{9}{11}\selectfont
\setlength{\tabcolsep}{3pt}
\caption{
Evaluation results of accent conversion.
Bold and underlined values mark the best and second-best eligible scores.
Gray cells are excluded from ranking for $^{\dagger}$using text input, and $^{\ddagger}$using reference speech.
$^{*}$Performance is a reproduced result.
}
\label{tab:main}
\begin{tabular*}{\textwidth}{@{\extracolsep{\fill}}llcccccccc@{}}
\toprule
Task & System & WER (\%) $\downarrow$ & Spk Sim $\uparrow$ & DNSMOS $\uparrow$ & $\Delta$ P(us) & Arrival (\%) $\uparrow$ & MOS-N & MOS-A & Bnd. Err. (ms) $\downarrow$ \\
\midrule

\multirow{5}{*}{Normalization} & \textbf{\our{}} & \underline{16.9} & 0.388 & \textbf{3.102} & \underline{0.424} & \textbf{84.4} & \underline{3.75} & \textbf{3.30} & \textbf{58.7} \\


 & TokAN~\cite{tokan} & 19.9 & \textbf{0.410} & 2.946 & 0.208 & \underline{77.4} & 3.43 & 2.61 & 407.0 \\
 
 & CosyAccent~\cite{cosyaccent} & \textbf{10.9} & \underline{0.397} & \underline{3.049} & 0.202 & 67.8 & 3.44 & 2.92 & \underline{218.2} \\
 

 & Vevo-Style$^{\ddagger}$~\cite{vevo} & 29.2 & \textcolor{gray}{0.657} & 2.993 & \textbf{0.433} & 54.8 & \textbf{3.91} & \underline{2.98} & 310.5 \\
 
 & RWA$^{\dagger*}$~\cite{rwa} & \textcolor{gray}{3.2} & 0.284 & 2.832 & 0.214 & 53.0 & 1.58 & 1.16 & 271.8 \\
\midrule

\multirow{4}{*}{L1$\to$L2} & \textbf{\our{}} & 17.1 & \underline{0.500} & \underline{3.189} & $-$0.207 & \textbf{13.3} & \underline{3.66} & \textbf{3.44} & \textbf{45.5} \\
 & TokAN L1$\to$L2$^{*}$~\cite{bagat} & \textbf{5.8} & \textbf{0.535} & \textbf{3.215} & $-$0.028 & 4.2 & \textbf{3.95} & 2.51 & 254.6 \\
 
 & Vevo-Style$^{\ddagger}$~\cite{vevo} & \underline{6.5} & \textcolor{gray}{0.685} & 3.167 & \textbf{$-$0.398} & 3.3 & \underline{3.66} & 3.09 & \underline{173.5} \\
 
 & RWA$^{\dagger*}$~\cite{rwa} & \textcolor{gray}{3.4} & 0.327 & 2.899 & \underline{$-$0.260} & \underline{6.7} & 1.08  & 1.22 & 219.7 \\
\bottomrule
\end{tabular*}
\end{table*}

\vspace{-0.1cm}
\subsection{Speech activity conditioning}

Matching the generation length to the source duration alone does not specify where speech and pauses should occur.
Inspired by dubbing methods~\cite{dubs2st}, we represent speech activity as a binary sequence $v_{1:T}$ on the output-unit timeline.
Each indicator is one for speech and zero for non-speech, as shown in Fig.~\ref{fig:architecture}.
We embed each indicator and add it to the masked unit feature, providing a frame-level speech--pause schedule.
During training, the activity sequence and output length are obtained from the target speech, and from the source during inference.
This allows the model to adjust unit identities to the requested accent while adhering to the source activity.

\section{Experimental Setup}
\label{sec:experiments}
\vspace{-0.1cm}
\subsection{Implementation details}
\label{sec:implementation}

For datasets, we construct natural pairs from CMU ARCTIC~\cite{arctic} and L2-ARCTIC~\cite{l2arctic}.
The datasets contain collectively 29,585 recordings from 27 speakers and 1,132 prompts, and the cross-pairing produces 588,714 pairs.
The pairs are then divided into train/val/test subsets in an 8:1:1 ratio, where validation and test sets include only normalization and L1-to-L2 conversions.
For pretraining, we use LibriTTS-R recordings~\cite{librittsr}.
For conversion training, we sample the training batch from the proposed dataset and the CosyAccent dataset~\cite {cosyaccent} with probabilities of 75\% and 25\%, respectively.

The encoder consists of Whisper-medium~\cite{whisper} and linear layer with LayerNorm as the adapter.
The diffusion transformer contains 8 blocks with 512 hidden size, 8 attention heads, and 2048 feed-forward dimension, comprising approximately 55.4M trainable parameters. 
We extract speech units from HuBERT-Large layer 17~\cite{hubert} and use k-means discretization with 1000 clusters.
Waveforms are synthesized with a pretrained unit-to-mel renderer~\cite{tokan} and BigVGAN vocoder~\cite{bigvgan}.
Following~\cite{tokan, cosyaccent, dlman}, we pretrain the converter for 24k updates with a batch size of 32.
We apply run-length augmentation with factors in $[0.9,1.1]$ with probability 0.5 and drop the speech activity condition with probability 0.1.
Training uses AdamW with $3\times10^{-4}$ learning rate, a 4k step warm-up, betas $(0.9,0.98)$, weight decay 0.01, and gradient clipping at 1.
We select the checkpoint with the best validation performance with 20 unmasking steps for inference.
All experiments were conducted on an NVIDIA RTX 4090 GPU.

\vspace{-0.1cm}
\subsection{Objective evaluation}
We report word error rate (WER) using Whisper-large-v3 for intelligibility.
Source-speaker cosine similarity is measured with ECAPA-TDNN~\cite{ecapa}, and autmoatic naturalness is obtained with DNSMOS~\cite{dnsmos835}.
Silero VAD\footnote{\url{https://github.com/snakers4/silero-vad}} (threshold 0.5; 20-ms grid) is used to calculate boundary error by averaging the source boundary’s distance to the nearest generated boundary.

Accent conversion performance is crucial for validating the proposed method, but no established metric accurately captures performance across multiple accents. 
To effectively measure accent conversion, we utilize CommonAccent XLSR~\cite{commonaccent}, an established accent classifier, and obtain the US-accent probability change from source \(x\) to output \(\hat{x}\): $\Delta P(\mathrm{us})=P(\mathrm{us}\mid\hat{x})-P(\mathrm{us}\mid x)$.
Positive/negative values indicate increased/decreased US probability.
We also formulate Accent Arrival, which compares de-duplicated speech units extracted from output and reference waveforms.
We score output \(u\) against the same-prompt reference \(r\) using longest common subsequence (LCS): $ \mathrm{LCSR}(u,r)=\mathrm{LCS}(u,r)/\max(|u|,|r|),$
where LCSR is the ratio of LCS.
We create accent groups using samples from up to three reference speakers, and measure the highest-scoring group as the assignment.
Accent Arrival is the percentage assigned to the requested group relative to eligible pairs.

\vspace{-0.1cm}
\subsection{Subjective evaluation}

The evaluation of the perceptual quality of the generated speech plays a key role, especially in how the accents are represented.
We conduct subjective listening tests with 20 professionals.
Each listener evaluates 20 randomly selected samples in total across normalization and L1-to-L2 conversion, with the presentation order randomized independently for each evaluation session.
Naturalness (MOS-N) is rated without a reference, and accented similarity (MOS-A) is rated by asking to hear the target recording as a reference and rate how closely the converted speech matches its accent.
Both dimensions use a five-point scale from 1 to 5, with higher scores indicating greater naturalness or closer accent similarity.

\begin{figure}[t]
\centering
\includegraphics[width=\columnwidth]{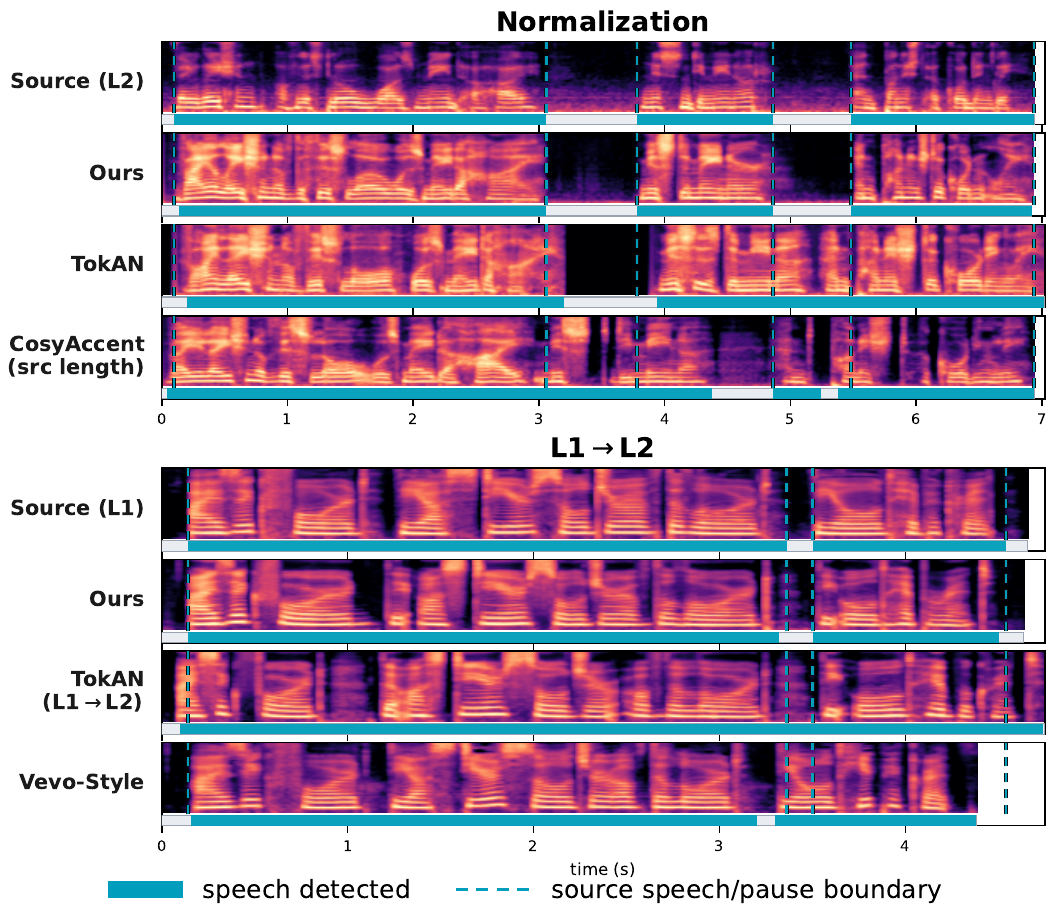}
\caption{
    Speech/pause timing for normalization (top) and L1-to-L2 conversion (bottom).
    Cyan bars indicate detected speech, and dashed lines mark source boundaries.
}
\label{fig:activity}

\end{figure}

\vspace{-0.1cm}
\section{Results}
\label{sec:results}
\vspace{-0.1cm}
\subsection{Main comparison}

Table~\ref{tab:main} shows that \our{} achieves the highest Accent Arrival and mean listener-rated accent similarity in both directions, providing complementary evidence of improved requested-accent matching.
For normalization, these accent gains produce the highest DNSMOS and second-highest mean MOS-N, suggesting that improved accent matching is compatible with natural-sounding speech.
Normalization WER falls between TokAN and CosyAccent, with source-speaker similarity close to both.
Thus, the shared bidirectional model remains competitive with systems designed specifically for normalization.

The distinction between changing the source accent and matching the requested accent is especially important for L1-to-L2 conversion.
Vevo-Style reduces US-accent probability more than \our{}, yet its Accent Arrival is lower.
A larger shift away from US-accent probability therefore does not necessarily indicate closer resemblance to the requested L2 accent.
Human evaluation supports this interpretation: \our{} also achieves the highest mean MOS-A, compared with Vevo-Style and TokAN.
Although the low absolute Arrival indicates the difficulty of the requested accent matching, \our{} outperforms the evaluated baselines.

These accent gains come with a clear WER trade-off.
For L1-to-L2 conversion, \our{} yields 17.1\% WER, compared with TokAN's 5.8\% and Vevo-Style's 6.5\%, despite competitive naturalness and speaker similarity.
The baselines' lower WER therefore does not coincide with better requested-accent matching.
WER alone does not establish accent fidelity~\cite{zhong2025pairwise}, so we interpret these outcomes jointly: stronger requested-accent matching is accompanied by higher recognition error.

\vspace{-0.15cm}
\subsection{Speech synchronization}
\vspace{-0.1cm}
Speech duration plays a crucial role in the perceived accent authenticity~\cite{onda2025duration}. Nevertheless, under duration and activity constraints, \our{} achieves higher accent conversion performance and competitive naturalness and speaker similarity, while reducing the boundary error by approximately 73\% over the best baseline.
Figure~\ref{fig:activity} shows that our outputs preserve source speech/pause placement in both scenarios.
TokAN and Vevo-Style generate speech that mismatches not only the alignment but also the total duration.
Although CosyAccent addresses dubbing with duration control, it misplaces internal pauses.
This provides valid evidence that \our{} produces synchronized speech suitable for dubbing, and shows that matching total duration alone is insufficient.

\begin{table}[t]
\centering
\fontsize{9}{11}\selectfont
\setlength{\tabcolsep}{2.5pt}
\caption{
Ablation results.
Slash separates results for normalization and L1-to-L2 conversion.
The last three rows vary the conversion-stage training data for the same model.
}
\label{tab:ablation}
\begin{tabular*}{\columnwidth}{@{\extracolsep{\fill}}lccc@{}}
\toprule
Variant & WER (\%) $\downarrow$ & Arrival (\%) $\uparrow$ & Bnd. Err. (ms) $\downarrow$ \\
\midrule
\our{} & \textbf{16.9} / 17.1 & 84.4 / 13.3 & 58.7 / 45.5 \\
w/o activity guide & 19.0 / 16.7 & \textbf{88.7} / 14.2 & 430 / 118 \\
\midrule[0.25pt]
ARCTIC cross & 20.9 / 23.5 & 83.5 / 17.5 & 64.1 / 39.0 \\
ARCTIC one-to-one & 31.5 / 26.4 & 73.0 / \textbf{21.7} & 75.1 / 46.7 \\
CosyAccent only & 17.7 / \textbf{8.51} & 75.7 / 1.67 & \textbf{23.5} / \textbf{27.8} \\
\bottomrule
\end{tabular*}
\vspace{-0.35cm}
\end{table}

\vspace{-0.15cm}
\subsection{Ablation study}
\vspace{-0.1cm}

Table~\ref{tab:ablation} evaluates speech activity conditioning and conversion training data for our proposed method.
Removing activity conditioning increases boundary error from 58.7 to 430 ms for normalization and from 45.5 to 118 ms for L1-to-L2 conversion.
Arrival increases in both directions, revealing a trade-off between temporal alignment and requested-accent assignment.
Cross-pairing lowers WER in both directions relative to one-to-one pairing.
Normalization Arrival rises from 73.0\% to 83.5\%, while L1-to-L2 Arrival decreases from 21.7\% to 17.5\%.
CosyAccent-only training yields lower WER than both real-only recipes, with Arrival reaching 75.7\% for normalization and only 1.67\% for L1-to-L2 conversion.
The mixed recipe achieves lower WER than both real-only recipes, at some cost to L1-to-L2 Arrival.
Relative to synthetic-only training, L1-to-L2 Arrival rises from 1.67\% to 13.3\% and WER from 8.51\% to 17.1\%.

\vspace{-0.2cm}
\section{Conclusion}
\vspace{-0.15cm}
\label{sec:conclusion}
We presented \our{}, combining real-speech cross-pairing with activity-guided discrete diffusion for bidirectional accent conversion.
The model improves accent matching and source synchronization over the compared baselines, supporting synchronized dubbing with competitive naturalness and speaker similarity.
The ablations show that intelligibility alone is insufficient: conversion for dubbing must also realize the requested accent and preserve source timing.

\bibliographystyle{IEEEbib}
\bibliography{shortstrings,refs}

@inproceedings{whisper,
  author = {Radford, Alec and Kim, Jong Wook and Xu, Tao and Brockman, Greg and McLeavey, Christine and Sutskever, Ilya},
  title = {Robust Speech Recognition via Large-Scale Weak Supervision},
  booktitle = icml,
  year = {2023}
}

@article{hubert,
  author = {Hsu, Wei-Ning and Bolte, Benjamin and Tsai, Yao-Hung Hubert and Lakhotia, Kushal and Salakhutdinov, Ruslan and Mohamed, Abdelrahman},
  title = {{HuBERT}: Self-Supervised Speech Representation Learning by Masked Prediction of Hidden Units},
  journal = taslp,
  year = {2021}
}

@inproceedings{tokan,
  author = {Bai, Qibing and Inoue, Sho and Wang, Shuai and Jiang, Zhongjie and Wang, Yannan and Li, Haizhou},
  title = {Accent Normalization Using Self-Supervised Discrete Tokens with Non-Parallel Data},
  booktitle = is,
  year = {2025}
}

@inproceedings{cosyaccent,
  author = {Bai, Qibing and Shi, Shuhao and Wang, Shuai and Ju, Yukai and Wang, Yannan and Li, Haizhou},
  title = {{CosyAccent}: Duration-Controllable Accent Normalization Using Source-Synthesis Training Data},
  booktitle = icassp,
  year = {2026}
}

@inproceedings{rwa,
  author = {Ezzerg, Abdelhamid and Merritt, Thomas and Yanagisawa, Kayoko and Bilinski, Piotr and Proszewska, Magdalena and Pokora, Kamil and Korzeniowski, Renard and Barra-Chicote, Roberto and Korzekwa, Daniel},
  title = {Remap, Warp and Attend: Non-Parallel Many-to-Many Accent Conversion with Normalizing Flows},
  booktitle = slt,
  year = {2023}
}

@inproceedings{vevo,
  author = {Zhang, Xueyao and Zhang, Xiaohui and Peng, Kainan and Tang, Zhenyu and Manohar, Vimal and Liu, Yingru and Hwang, Jeff and Li, Dangna and Wang, Yuhao and Chan, Julian and Huang, Yuan and Wu, Zhizheng and Ma, Mingbo},
  title = {Vevo: Controllable Zero-Shot Voice Imitation with Self-Supervised Disentanglement},
  booktitle = iclr,
  year = {2025}
}

@inproceedings{arctic,
  author = {Kominek, John and Black, Alan W.},
  title = {The {CMU Arctic} Speech Databases},
  booktitle = ssw,
  year = {2004}
}

@inproceedings{l2arctic,
  author = {Zhao, Guanlong and Sonsaat, Sinem and Silpachai, Alif and Lucic, Ivana and Chukharev-Hudilainen, Evgeny and Levis, John and Gutierrez-Osuna, Ricardo},
  title = {{L2-ARCTIC}: A Non-native English Speech Corpus},
  booktitle = is,
  year = {2018}
}

@inproceedings{dlman,
  author = {Bai, Qibing and Du, Yuhan and Ko, Tom and Wang, Shuai and Wang, Yannan and Li, Haizhou},
  title = {Controllable Accent Normalization via Discrete Diffusion},
  booktitle = is,
  year = {2026}
}

@inproceedings{bagat,
  author = {Bagat, Rapha{\"e}l and Zhang, Zhe and Yamagishi, Junichi and Illina, Irina and Vincent, Emmanuel},
  title = {Synthetic Audio Generation Framework for Air Traffic Control Speech Recognition},
  booktitle = is,
  year = {2026}
}

@inproceedings{bigvgan,
  author = {Lee, Sang-gil and Ping, Wei and Ginsburg, Boris and Catanzaro, Bryan and Yoon, Sungroh},
  title = {{BigVGAN}: A Universal Neural Vocoder with Large-Scale Training},
  booktitle = iclr,
  year = {2023}
}

@inproceedings{librittsr,
  author = {Koizumi, Yuma and Zen, Heiga and Karita, Shigeki and Ding, Yifan and Yatabe, Kohei and Morioka, Nobuyuki and Bacchiani, Michiel and Zhang, Yu and Han, Wei and Bapna, Ankur},
  title = {{LibriTTS-R}: A Restored Multi-Speaker Text-to-Speech Corpus},
  booktitle = is,
  year = {2023}
}

@inproceedings{jin2023multiple,
  author = {Jin, Mumin and Serai, Prashant and Wu, Jilong and Tjandra, Andros and Manohar, Vimal and He, Qing},
  title = {Voice-Preserving Zero-Shot Multiple Accent Conversion},
  booktitle =icassp,
  year = {2023}
}

@inproceedings{jia2023psdn,
  author = {Jia, Dongya and Tian, Qiao and Peng, Kainan and Li, Jiaxin and Chen, Yuanzhe and Ma, Mingbo and Wang, Yuping and Wang, Yuxuan},
  title = {Zero-Shot Accent Conversion using Pseudo Siamese Disentanglement Network},
  booktitle = is,
  year = {2023}
}

@inproceedings{inoue2025macst,
  author = {Inoue, Sho and Wang, Shuai and Wang, Wanxing and Zhu, Pengcheng and Bi, Mengxiao and Li, Haizhou},
  title = {{MacST}: Multi-Accent Speech Synthesis via Text Transliteration for Accent Conversion},
  booktitle = {Proc. ICASSP},
  year = {2025}
}

@inproceedings{nguyen2025distillation,
  author = {Nguyen, Tuan Nam and Akti, Seymanur and Pham, Ngoc Quan and Waibel, Alexander},
  title = {Improving Pronunciation and Accent Conversion through Knowledge Distillation And Synthetic Ground-Truth from Native {TTS}},
  booktitle = {Proc. ICASSP},
  year = {2025}
}

@inproceedings{dubs2st,
  author = {Choi, Jeongsoo and Kim, Jaehun and Chung, Joon Son},
  title = {{Dub-S2ST}: Textless Speech-to-Speech Translation for Seamless Dubbing},
  booktitle = {Findings of EMNLP},
  year = {2025}
}

@inproceedings{onda2025duration,
  author = {Onda, Kentaro and Imoto, Keisuke and Fukayama, Satoru and Saito, Daisuke and Minematsu, Nobuaki},
  title = {Prosodically Enhanced Foreign Accent Simulation by Discrete Token-based Resynthesis Only with Native Speech Corpora},
  booktitle = {Proc. Interspeech},
  year = {2025}
}

@inproceedings{mdlm,
  author = {Sahoo, Subham Sekhar and Arriola, Marianne and Schiff, Yair and Gokaslan, Aaron and Marroquin, Edgar and Chiu, Justin T and Rush, Alexander and Kuleshov, Volodymyr},
  title = {Simple and Effective Masked Diffusion Language Models},
  booktitle = {Proc. NeurIPS},
  year = {2024}
}

@inproceedings{maskgit,
  author = {Chang, Huiwen and Zhang, Han and Jiang, Lu and Liu, Ce and Freeman, William T.},
  title = {{MaskGIT}: Masked Generative Image Transformer},
  booktitle = {Proc. CVPR},
  year = {2022}
}

@inproceedings{zhong2025pairwise,
  author = {Zhong, Jinzuomu and Liu, Suyuan and Wells, Dan and Richmond, Korin},
  title = {Pairwise Evaluation of Accent Similarity in Speech Synthesis},
  booktitle = {Proc. Interspeech},
  year = {2025}
}

@inproceedings{commonaccent,
  author = {Zuluaga-Gomez, Juan and Ahmed, Sara and Visockas, Danielius and Subakan, Cem},
  title = {{CommonAccent}: Exploring Large Acoustic Pretrained Models for Accent Classification Based on Common Voice},
  booktitle = {Proc. Interspeech},
  year = {2023},
  pages = {5291--5295},
  doi = {10.21437/Interspeech.2023-2419}
}

@inproceedings{ecapa,
  author = {Desplanques, Brecht and Thienpondt, Jenthe and Demuynck, Kris},
  title = {{ECAPA-TDNN}: Emphasized Channel Attention, Propagation and Aggregation in {TDNN} Based Speaker Verification},
  booktitle = is,
  year = {2020},
}

@inproceedings{dnsmos835,
  author = {Reddy, Chandan K. A. and Gopal, Vishak and Cutler, Ross},
  title = {{DNSMOS P.835}: A Non-Intrusive Perceptual Objective Speech Quality Metric to Evaluate Noise Suppressors},
  booktitle = icassp,
  year = {2022},
}

@string{cvpr   = "Proc. CVPR"}

@string{emnlp  = "Proc. EMNLP"}

@string{icassp="Proc. ICASSP"}

@string{iclr = "Proc. ICLR"}

@string{icml =  "Proc. ICML"}

@string{is =  "Proc. Interspeech"}

@string{neurips = nips}

@string{slt = "Proc. SLT"}

@string{ssw = "Proc. ISCA Speech Synthesis Workshop"}

@string{taslp =  "IEEE/ACM Trans. on Audio, Speech, and Language Processing"}
\end{document}